\documentclass[pre,aps,floatfix,twocolumn,superscriptaddress,10pt]{revtex4-2}

\usepackage{amsmath,amssymb}
\usepackage{graphics,graphicx}
\usepackage{hyperref}
\hypersetup{colorlinks=true, citecolor=blue, urlcolor=blue, linkcolor=blue}

\begin{document}

\title{Record statistics based prediction of fracture in the random spring network model}

\author{Subrat Senapati}
\email{subrat.senapati52@gmail.com}
\affiliation{Department of Applied Mechanics, Indian Institute of Technology Madras, Chennai-600036, India}
\author{Subhadeep Roy}
\email{subhadeep.r@hyderabad.bits-pilani.ac.in}
\affiliation{Department of Physics, Birla Institute of Technology \& Science Pilani, Hyderabad Campus, Secunderabad, Telangana 500078, India}
\affiliation{The Institute of Mathematical Sciences, C.I.T. Campus, Taramani, Chennai-600113, India} 
\author{Anuradha Banerjee}
\email{anuban@iitm.ac.in}
\affiliation{Department of Applied Mechanics, Indian Institute of Technology Madras, Chennai-600036, India}
\author{R.Rajesh} 
\email{rrajesh@imsc.res.in}
\affiliation{The Institute of Mathematical Sciences, C.I.T. Campus, Taramani, Chennai-600113, India} 
\affiliation{Homi Bhabha National Institute, Training School Complex, Anushakti Nagar, Mumbai-400094, India}

\date{\today}

\begin{abstract}
We study the role of record statistics of damage avalanches in predicting the fracture of a heterogeneous material under tensile loading. The material is modeled using a two-dimensional random spring network where disorder is introduced through randomness in the breakage threshold strains of the springs. It is shown that the waiting time between successive records of avalanches has a maximum for moderate disorder, thus showing an acceleration of records with impending fracture. Such a signature is absent for low disorder strength when the fracture is nucleation-dominated, and high disorder strength when the fracture is percolation type. We examine the correlation between the record with the maximum waiting time and the crossover record at which the avalanche statistics change from off-critical to critical. Compared to the avalanche based predictor for failure, we show that the record statistics have the advantage of both being real-time as well as able to predict final fracture at much smaller strains. We also show that in the avalanche-dominated regime, the failure strain is shown to have a linear relation with the strain at the maximum waiting time, making possible a quantitative prediction.  
\end{abstract}

\maketitle

\section{Introduction}

The fracture of materials with micro and mesoscale heterogeneity is  accompanied by crackling noise~\cite{herrmann2014statistical,chakrabarti1997statistical}, i.e., the final fracture is preceded by intermittent bursts or avalanches of micro-cracking which generate acoustic emissions~\cite{petri1994experimental,baro2013statistical,salje2014crackling,rosti2009crackling}. Experimental and theoretical studies have revealed that the intensity of the precursor activity depends on the degree of material disorder. In the limiting case of zero disorder, the ultimate failure occurs abruptly with hardly any precursors~\cite{zapperi1997first,menezes2010influence}. However, at higher disorder, a gradual accumulation of damage is observed with an increasing rate of breaking bursts as failure is approached~\cite{ramos2013experimental,sornette2002predictability}. In materials like porous glass, it is possible to observe the role of disorder on damage accumulation by controlling the degree of heterogeneity~\cite{vasseur2015heterogeneity}.  Other instances of occurrence of avalanches can be found in the failure of biological material~\cite{garcimartin1997statistical,baro2016avalanche}, and construction material~\cite{petri1994experimental}, during creep in cellular glass~\cite{maes1998criticality}, in hydrogen precipitation in niobium~\cite{cannelli1993self}, in dislocation motion in ice crystals~\cite{weiss1997acoustic}, and  volcanic activity~\cite{diodati1991acoustic}. These phenomena are also reminiscent of the Gutenberg-Richter law for earthquake statistics~\cite{gutenberg1944frequency}. 

It has become apparent that avalanche dominated response is the rule rather than the exception in driven disordered systems. Examples outside fracture include the motion of domain walls in magnets (the Barkhausen effect)~\cite{zapperi1998dynamics} and flux lines in superconductors~\cite{field1995superconducting}, frictional sliding~\cite{ciliberto1994experimental},  fluid flow in porous media~\cite{thompson1987mercury} and the inflation of degassed lungs~\cite{suki1994avalanches}. Thus, identifying the statistical signatures of avalanche dynamics and the associated physical mechanisms responsible for them go well beyond the study of breakdown and fracture.   

To quantify the avalanche dynamics, the  tails of the probability distribution of the avalanche sizes have been characterized for fracture in different materials. In most cases, the distribution $P(s)$ of avalanche $s$ follows a power law distribution:
\begin{align}
\label{eq1}
P(s) \sim s^{-\eta}, \quad s \to \infty,
\end{align}    
where $\eta$ appears to depend on the nature of the material under tension/compression. Experimental values of $\eta$ obtained from fracture testing are 1.95 for cellular glass~\cite{maes1998criticality}, 2.0 for volcanic rocks~\cite{diodati1991acoustic}, 0.85 for hydrogen precipitation~\cite{cannelli1993self}, 0.52--0.84 for concrete~\cite{petri1994experimental}, 2.31--3.59 for porous stainless steel under tension~\cite{chen2020avalanches}, 1.7--2.5 for granular Mg-Ho alloys under compression and tension~\cite{chen2019acoustic}, 1.3--1.4 for porcine bone~\cite{baro2016avalanche}, etc. 

In addition to experiments, the understanding of the breakdown of disordered systems has progressed to a large extent with the use of large-scale simulations of discrete models~\cite{herrmann2014statistical}. For instance, a conductor is represented by a resistor network or an elastic continuum by a network of springs or beams, or they could be simpler models like the fiber bundle model which while not representing the continuum are more analytically tractable.  The disorder is usually modeled by random failure thresholds or elastic heterogeneity. In the random resistor/fuse model, the avalanche exponent depends on the lattice structure, it being 2.75  for a diamond lattice, 3.05 for a triangular lattice~\cite{zapperi2005crack}, and close to 2.5 in three dimensions~\cite{zapperi2005crack3d}, 2.5 being the mean field exponent observed in the fiber bundle model~\cite{hemmer1992distribution,hansen1994criticality,hansen2015fiber}. More recent studies show that the exponents also depend on the extent of disorder~\cite{shekhawat2013damage,moreira2012fracturing}. In the random spring network model, the avalanche exponent in two dimensions was found to be close to the mean field result of 2.5~\cite{zapperi1997first}. This result is not affected when hardening is included~\cite{kumar2022interplay}. However, the value is found to be $1.8$--$2.0$ with elastic heterogeneity as found in bone~\cite{mayya2016splitting,mayya2017role,mayya2018role}, close to 1.3 in the presence of a crack~\cite{parihar2020role} and varying with composition for two phase materials~\cite{senapati2023role}.

In addition to looking for scale free behavior and universality in avalanche distribution, effort has gone in using avalanche statistics to predict  imminent failure. In particular, avalanche statistics are not stationary, and carry a signature of acceleration before final failure, indicated by increase in seismic or acoustic signals, rate of deformation, etc~\cite{nataf2014predicting,hao2013power,vasseur2015heterogeneity,xu2019criticality,main1999applicability,bell2011challenges,michlmayr2017fiber,petri1994experimental,baro2018experimental,sammonds1992role}. The strain-dependent avalanche distribution is seen to have a crossover from one power-law to another power-law with a smaller exponent as strain approaches the failure strain. The latter avalanches are referred to as critical avalanches. In the fiber bundle model, the exponent changes from $5/2$ to $3/2$ near breakdown, while in the two dimensional random resistor network,  the exponent decreases from 3 to 2 near catastrophic failure~\cite{pradhan2005crossover,pradhan2005crossover}.  Experimentally, the energy avalanche exponent was seen to decrease from $1.7$ to $1.55$ for sandstone and from $1.5$ to $1.3$ for coal~\cite{jiang2016collapsing}, wherein the critical avalanches were shown to be localized compared to the initial avalanches. We note that this decrease in exponent is equivalent to the decrease of the {\it b-value} while approaching a main shock during a seismic event~\cite{imoto1991changes,nanjo2012decade} or even in laboratory experiments like rock fracture~\cite{scholz1968frequency}. The change in exponent has been suggested as a predictor or indicator of imminent failure.

Another  precursor that has been suggested more recently as a predictor of failure is based on record statistics of avalanches. A record avalanche is an avalanche that is  bigger than all previous avalanches. The acceleration of activity close to failure shows up as more records being rapidly created. The waiting time between successive records was shown to first increase, reach a maximum at the $k^{\ast}$-th record and then decrease, based on simulations of a model for porous granular material~\cite{pal2016record} and the mean field fiber bundle model in the quasi-brittle regime~\cite{kadar2020record}. This feature was also shown to be present in the energy avalanches of the local load sharing fiber bundle model~\cite{roy2023record}. It was shown that the characteristic index $k^{\ast}$ correlates with the record index at which the avalanche exponent changed from off-critical to critical in failure of coal under compression~\cite{jiang2017predicting}. However, the predictor based on analysis of record statistics has an advantage over the predictor based on crossover in the avalanche distribution in that real-time monitoring of records is possible. 

In this paper, we focus on record statistics of fracture of the random spring network model (RSNM) with disordered strain-based threshold for rupture. Compared to the fiber bundle model, RSNM is more realistic and captures local stress concentration as well as interactions between defects, and has the advantage that it captures the continuum elastic behavior at the macroscopic scale. In addition, it distinguishes between failure under compression and tension, unlike other discrete models. RSNM has been effective in reproducing several features of experimental fracture data of heterogeneous media, such as power-law statics of cracking  events~\cite{zapperi1997first,kumar2022interplay,mayya2016splitting,mayya2017role}, quasi-brittle macroscopic response resulting from inherent disorder~\cite{curtin1990mechanics,urabe2010fracture,senapati2023role,bolander2005irregular,yip2006irregular,wang2020lattice,suryawanshi2023novel}, complex failure paths~\cite{zapperi1997first,kumar2022interplay}, effect of patterning~\cite{dimas2014coupled,dimas2015random,parihar2020role,boyina2015mixed}, etc. To study the dependence of record statistics on disorder, we vary the extent of disorder from high to low, corresponding to the fracture type changing from percolation to avalanche-dominated to nucleation-type, as demonstrated for the random resistor network in Ref.~\cite{shekhawat2013damage}. We show that the waiting times between records shows a maximum at a non-trivial $k^\ast$-th record when the fracture is avalanche dominated. For percolation-type fracture, there is only one record while for nucleation-type, the waiting times decrease with increasing number of records.  We show that, in the avalanche-dominated regime, there is a linear relation between the failure strain and strain at maximum stress with the strain at the $k^\ast$-th record, or equivalently a quantitative prediction is possible. We also show a correlation between $k^\ast$, and the cross-over record at which off-critical avalanches crossover to critical avalanches. However, we find that while the predictability from the off critical--critical crossover becomes weaker with increasing system size, the predictor from records still remains effective.

\section{\label{sec:model}Model}

\begin{figure}
\includegraphics[width=\columnwidth]{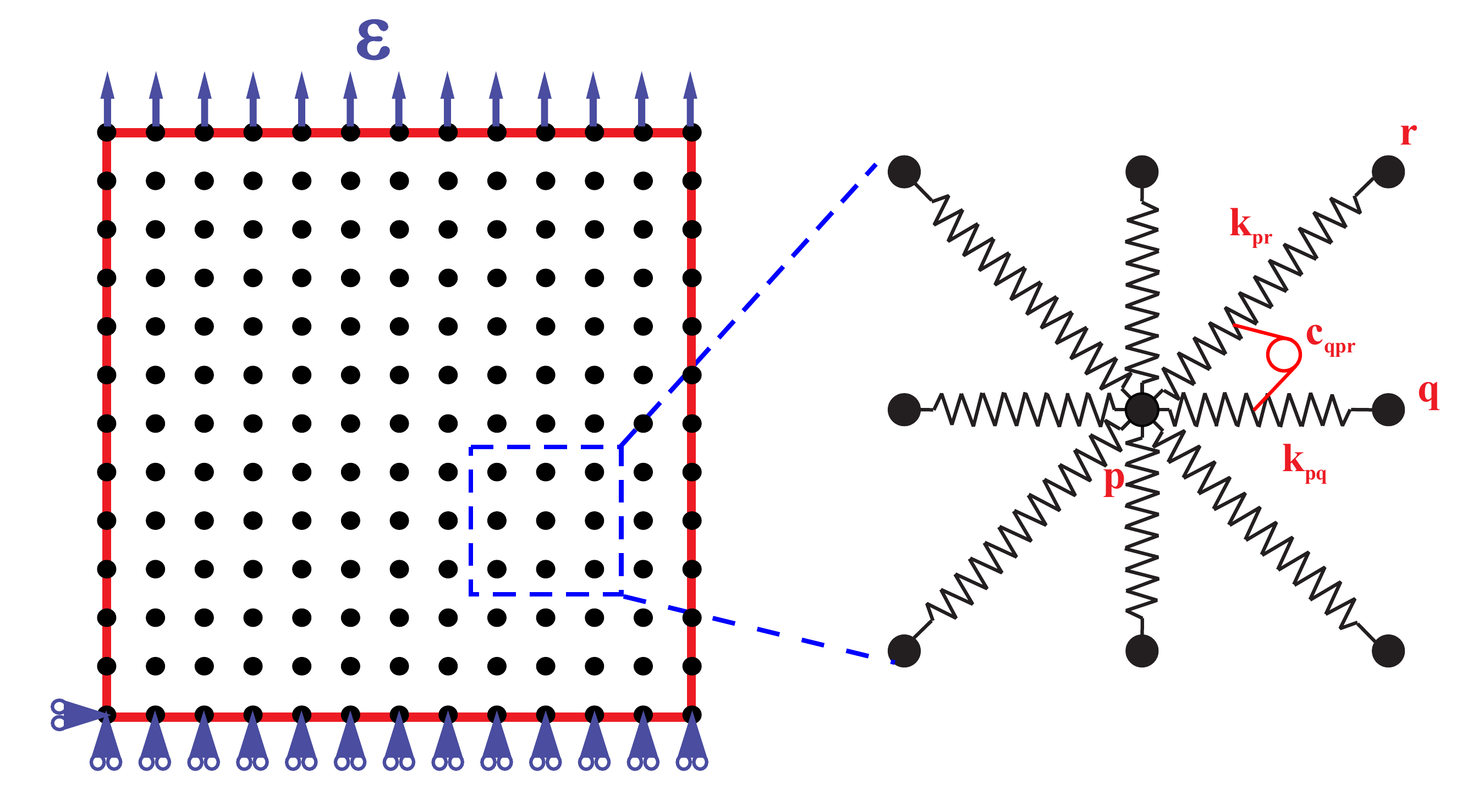}
\caption{(a) The discretization and boundary conditions of a continuum domain. (b) The connectivity of a lattice site with its neighbors.}
\label{fig0}
\end{figure}

For simulation of fracture in heterogeneous materials we use the random spring network model (RSNM). To develop the network, we discretize the domain with a square lattice having lattice size, $a$. Each lattice point of the network, or equivalently particle,  is connected with its nearest neighbors and with its next neighbors through extensional springs. In addition, it interacts rotationally with adjacent pair of neighbors by torsional springs, as shown in Fig.~\ref{fig0}.

When the network is strained, the net potential energy $\Phi$ of the system is stored in both the extensional as well as torsional springs, denoted by $\Phi_{ext}$ and $\Phi_{tor}$, respectively:
\begin{equation}
\label{eq2}
\Phi = \Phi_{ext}+\Phi_{tor}.
\end{equation}    
The potential energy stored in extensional springs can be expressed as
\begin{equation}
\label{eq3}
\Phi_{ext} = \sum_{\left\langle ij\right\rangle}\frac{1}{2}k_{ij}\left(|\vec{r}_{i}-\vec{r}_{j}|-a_{ij}\right)^2,
\end{equation}
where the sum is over all pairs of particles connected by extensional springs, $a_{ij}$ denotes the undeformed distance between lattice points $i$ and $j$, $\vec{r}_{i}$ and $\vec{r}_{j}$ are their corresponding current position vectors and $k_{ij}$ is the stiffness of the extensional spring connecting them. The potential energy stored in the torsional springs is calculated as:
\begin{equation}
\label{eq4}
\Phi_{tor} = \sum_{\left\langle pqr\right\rangle}\frac{1}{2}c_{pqr}\left(\theta_{pqr}-\frac{\pi}{4}\right)^2,  
\end{equation}
where the sum is over all triplets of particles  connected by torsional springs, $\theta_{pqr}$ denotes the current angle subtended by two adjacent neighbors with the lattice point $p$ and $c_{pqr}$ denotes the torsional stiffness of the rotational spring.

The extensional and torsional stiffness of the respective springs can be expressed in terms of $a$ and the continuum elastic properties, Young's modulus, $E$, and Poisson's ratio, $\nu$, by applying equivalence of the strain energy density of the continuum with the  potential energy density of the network~\cite{monette1994elastic}. For a homogeneous isotropic domain, the elastic constants are expressed as:
\begin{eqnarray}
E &=& \frac{8k\left(k+\frac{c}{a^2}\right)}{3k+\frac{c}{a^2}},
\label{eq5} \\
\nu &=& \frac{\left(k-\frac{c}{a^2}\right)}{3k+\frac{c}{a^2}},
\label{eq6}
\end{eqnarray}
where $k$ is the spring constant of diagonal springs and $c_{pqr}=c$. The resulting spring constant of both horizontal and vertical springs would be $2 k$.

Uniaxial tensile strain is applied to the top row of the spring network while the bottom row is restrained to move only in the horizontal direction, as shown in Fig.~\ref{fig0}. The strain is applied at $0.0002$ per increment, and for every increment, the system is equilibrated by evolving the positions of the particles using Newton's laws of motion. A dissipative force term $-\gamma\vec{v}_p$ is included for convergence to equilibrium. The damping coefficient $\gamma$ for the dissipative force is set to be $0.8$ per unit time  to prevent excessive oscillations. The resultant force on any particle $p$ is computed as:
\begin{equation}
\label{eq7} 
\vec{a}_{p}= -\nabla_{\vec{r}_p}\phi,
\end{equation}  
where the mass is set to unity. In each loading step, the system evolves iteratively to attain the state of  quasi-static equilibrium and in each iterative step, the updated position vector $\vec{r}_p(t + \Delta t)$ of the lattice points are computed based on the position vector of last two time steps $\vec{r}_p(t)$ and $\vec{r}_p(t - \Delta t)$ using Verlet algorithm \cite{verlet1967computer} as:
\begin{equation}
\label{eq8} 
\vec{r}_p(t \!+ \!\Delta t) = \vec{r}_p(t)\left(2\!-\!\gamma\Delta t\right)-\vec{r}_p(t - \Delta t)\left(1\!-\!\gamma\Delta t\right)+ \vec{a}_p (\Delta t)^2,
\end{equation} 
where the velocity in the dissipative term is calculated using backward difference formula $\vec{v}_p(t)=[r(t)-r(t-\Delta t)]/\Delta t +O(\Delta t)$.

The system is assumed to be statically equilibrated once the kinetic energy of each particle is below a specified limit. Further, we verify that in equilibrium the forces in the top and bottom edges are equal within a pre-defined tolerance. If any of the surviving springs fail after equilibration, based on the breaking rule discussed below, then the system is again re-equilibrated following the same iterative process until no further breakage of springs occurs for the given applied strain. We note that when a spring breaks, the torsional springs associated with it are also considered broken.

We choose Young's modulus $E =  200GPa$, $\nu=0.3$, and set $a=0.5$mm. The failure strain threshold, $\epsilon_{f}$, of each spring is chosen independently from a distribution  characterized by a parameter $\beta$. If the strain in a spring exceeds its strain threshold, then it is considered to be broken. Following Ref.~\cite{shekhawat2013damage}, we choose the cumulative distribution function, $F(\epsilon_{f})$, to be
\begin{equation}
\label{eq9} 
F(\epsilon_{f}) = \epsilon_{f}^{\beta}, \quad 0\leq \epsilon_{f} \leq 1.
\end{equation}
The parameter $\beta$ controls the extent of disorder: $\beta\rightarrow 0$ corresponds to an infinite disordered system while $\beta\rightarrow\infty$ corresponds to a system having minimal disorder~\cite{shekhawat2013damage}.   In this paper, we investigate the fracture behavior for a wide spectrum of disorder, represented by $\beta = 0.1, 1.0,3.0,5.0$. 

\section{Results}

We perform simulations of the RSNM in two dimensions subjected to a tensile remotely applied strain. The two controlling parameters are the extent of disorder characterized by exponent $\beta$, as defined in Eq.~(\ref{eq9}), and the system size $L$. These two parameters are known to affect the type of failure in random resistor network~\cite{shekhawat2013damage,moreira2012fracturing}. It was shown that increasing $\beta$ (decreasing extent of disorder) or increasing $L$  changes the type of fracture from percolation dominated to avalanche dominated to nucleation driven. For the present article, we concentrate on the avalanche statistics during the failure process and in particular the record-breaking avalanches. To achieve variable extent of disorder and system sizes,  we simulate $\beta=0.1, 1.0, 3.0, 5.0$, for system sizes, $L= 50, 100, 200$.

\begin{figure}
\includegraphics[width=\columnwidth]{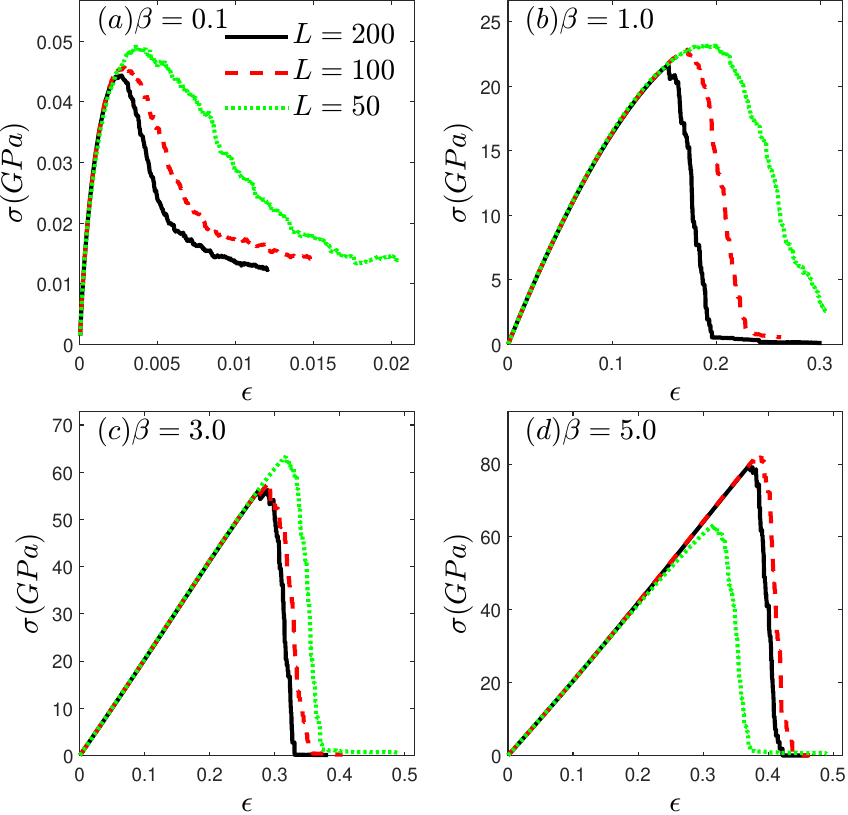}
\caption{Macroscopic stress-strain response for (a) $\beta=0.1$, (b) $\beta=1.0$, (c) $\beta=3.0$, and (d) $\beta=5.0$, where the thickness is taken to be 1 mm. The data are shown for different system sizes.}
\label{fig:macroscopic}
\end{figure}

The averaged macroscopic stress-strain responses for $\beta=0.1$, $1.0$, $3.0$, and $5.0$  are shown in Fig.~\ref{fig:macroscopic} for different system sizes. The peak stress for $\beta=0.1$ is about 500 times smaller than that for  $\beta\geq 1$. For $\beta=0.1$ there are large clusters of weak bonds that break for very small strains, resulting in the fracture being percolation-type as observed earlier for the random fuse model~\cite{shekhawat2013damage}.  For $\beta\geq 1$, the response is initially linear, followed by non-linear strain hardening at larger strains. The response can be seen to become more brittle with increasing $\beta$, as well as increasing $L$, when the fracture is expected to be more nucleation-type. The peak stress as well as failure strain increase with $\beta$. We find that for $\beta=0.1$, the first avalanche is the largest (also see later), and therefore we do the record analysis only for $\beta=1.0, 3.0, 5.0$.

\begin{figure}
\includegraphics[width=\columnwidth]{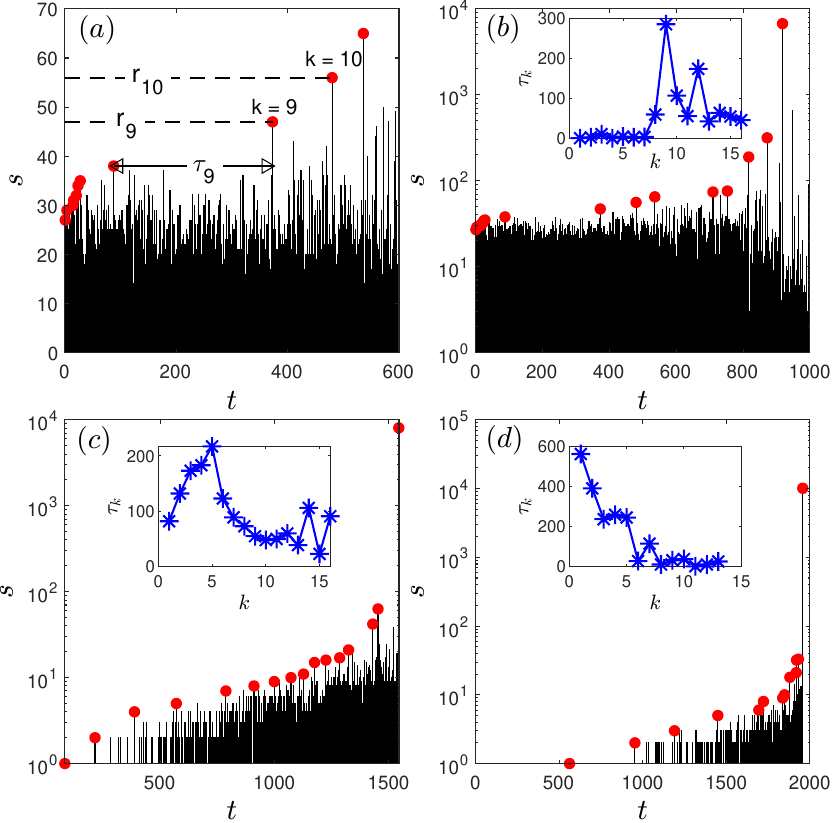}
\caption{(a) A window of a typical time series of avalanches (size denoted by $s$)  under uniaxial loading for $\beta=1.0$ and $L=200$. The vertical bars correspond to size of  avalanches  while the red dots represent record events. A record event is defined as an avalanche whose size is larger than all preceding avalanches. $r_k$ is the magnitudes of $k$-th record event while $\tau_k$ is the waiting time between the $(k-1)$-th  and $k$-th records.  (b), (c) and (d) show the full avalanche spectrum for $\beta=1.0$, $3.0$ and $5.0$ respectively for fixed system size $L=200$. The insets of (b), (c) and (d) show the variation of waiting time $\tau_k$ with the index $k$ of record events.  }
\label{fig1}
\end{figure}

Figure~\ref{fig1} shows the avalanches that we observe in the course of the failure process of a typical realization of uniaxial loading of the network. An avalanche is defined as the number of springs that break for an increment in applied strain. Figure~\ref{fig1}(a) shows the avalanche size $s$ as a function of time $t$ for $\beta=1.0$ and $L=200$. Time in our case is defined as follows: if at a certain instant, $d$ is the extension of the network over the course of the time evolution with $m$ consecutive steps of increment $\Delta d$ then the time required to achieve this extension $d$ will be $(d/\Delta d)t_0$, where $t_0$ is an internal time scale. We set $t_0=1$ without lose of generality. The vertical lines in Fig.~\ref{fig1}(a) represent all the observed avalanches within a time window while the red dots denote all the record-breaking avalanches. A record-breaking avalanche is defined as an avalanche whose size is larger than the previous record avalanche. For example, if we have a time sequence with avalanche sizes \{3, 1, 6, 8, 5, 7, 11\} then the sequence of the record-breaking bursts will be \{3, 6, 8, 11\}. 

Two quantities that can be associated with the record statistics are the size of the $k$-th record, $r_k$ and the waiting time. The waiting time $\tau_k$ is defined as the time elapsed between the  $k$-th record and the $(k-1)$-th record: 
\begin{align}
\tau_k = t_{k} - t_{k-1},
\end{align}       
where $t_{k}$ and $t_{k-1}$ are the times at which the record events $r_{k}$ and $r_{k-1}$ take place, and $t_0=0$. $\tau_k$ can also be interpreted as the lifetime of the $(k-1)$-th record. For example, in Fig.~\ref{fig1}(a), $\tau_9$ denotes the lifetime of the $8$-th record.

Figure~\ref{fig1}(b), (c) and (d)  show the full time series of avalanches for $\beta=1.0, 3.0, 5.0$ respectively for $L=200$. For $\beta=1$, in the time series shown in Fig.~\ref{fig1}(b), the onset of avalanches in contiguous increments is seen from the beginning of the simulation. Initially, a stationary region is observed in which the resulting record avalanches are seen to occur with increasing duration between them. As the failure process approaches final failure, the duration between two consecutive record avalanches starts decreasing. For comparatively less disorder,  $\beta=3.0$ and $5.0$, the initial stage has sparse population of avalanches. As the avalanches start to form frequently the resulting record avalanches appear to occur rapidly till final failure. Such trends and correlations in burst sequences can be analyzed by the statistics lifetime/waiting time of records, and by observing how it evolves with increasing rank $k$ of record events, at different extent of disorder. The insets of Fig.~\ref{fig1}(b), (c) and (d) show the variation of waiting time $\tau_k$ with rank $k$. For moderate $\beta$ (= 1.0, 3.0), $\tau_k$ is non-monotonic with a peak at a certain $k$ while such non-monotonic behavior is not seen for larger $\beta$. We will argue next that this peak has a signature of an upcoming catastrophic failure and can be correlated with the critical/failure strain.

\begin{figure}
\includegraphics[width=\columnwidth]{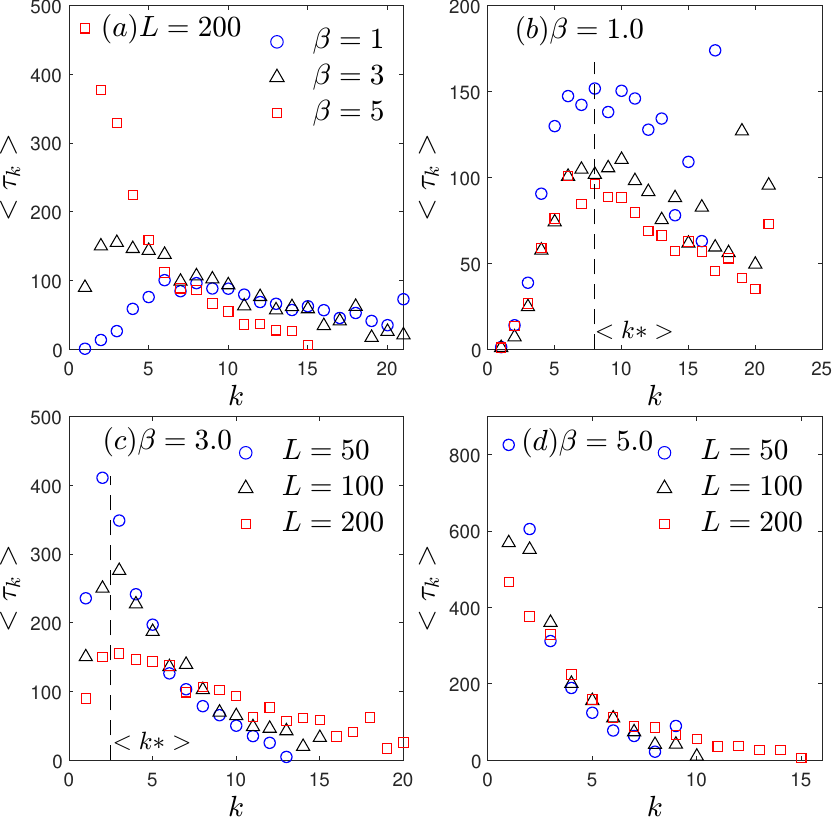}
\caption{Variation of the average waiting time,  $\langle\tau_k\rangle$, of the $k$-th record with index, $k$. (a) The data for different $\beta$ are compared for fixed system size $L=200$. The dependence of $\langle \tau_k \rangle$ on $L$ is shown for  (b) $\beta=1.0$, (c) $\beta=3.0$,  and (d) $\beta=5.0$. The location of the maximum of the curve, $\langle k^{\ast} \rangle$, appears to be independent of $L$ but shifts to a lower value as $\beta$ increases. } 
\label{fig2}
\end{figure}

To understand the effect of disorder on the waiting time series observed during the failure process, we study the variation of the average waiting time, $\langle \tau_k \rangle$, with $k$ for $L=200$ and $\beta=1.0$, 3.0 and 5.0, as presented in Fig.~\ref{fig2}(a). For high disorder, $\beta=1$, non-monotonic behavior is observed as the waiting time initially increases with $k$, then reaches a peak and decreases for larger $k$. Initially,  $\langle \tau_k \rangle$ is small, which corresponds to the initial rapid occurrences of record breaking events, then  $\langle \tau_k \rangle$ increases and reaches its peak which corresponds to the stationary region of the avalanche series seen earlier in Fig.~\ref{fig1}(b). The decrease seen for larger $k$ is a consequence of the acceleration in occurrence of record avalanches as the final failure event is approached.  We denote the rank $k$ at which $\langle \tau_k \rangle$ is maximum as $k^{\ast}$.  For lower disorder, $\beta=3.0$, the behavior of $\langle\tau_k\rangle$ is qualitatively similar to that of $\beta=1.0$, however the peak occurs at a lower rank $k$ and the waiting time for the initial records is higher comparatively. For $\beta=5.0$, when the disorder is even lower, the behavior is largely monotonic with initial record avalanches (at lower $k$) having the largest waiting time intervals and with increasing $k$ the waiting time decreasing monotonically.

Systems with lower disorder are known to exhibit nucleation type of fracture where the damage nucleates at only few locations and subsequent damage localizes in the neighborhood of the nucleated cracks which grow and interact until  a critical size is reached before final failure. Our observations are consistent with this understanding as the waiting time for nucleation are the largest for lower disorder ($\beta=5.0$) and there is no stationarity observed (see Fig.~\ref{fig1}~(d)) in the avalanche size time series as the fracture mechanisms become localized after nucleation.

We now examine the system size dependence of the lifetime statistics of records. For $\beta=1.0$, we find that $k^{\ast}$ does not change with $L$, however, the magnitude of $\langle \tau_k \rangle$ decreases with $L$ for all $k$, as seen in Fig.~\ref{fig2}(b). Similar features are seen for $\beta=3.0$ [see Fig.~\ref{fig2}(c)]. For $\beta=5.0$, $k^{\ast}=1$ for all $L$, as shown in Fig.~\ref{fig2}(d), and $\langle \tau_k \rangle$ decreases with $k$ for all $k$. Based on type of fracture, we expect that as $L$ is increased keeping $\beta$ fixed, the fracture type changes from avalanche type to nucleation type. Thus, we would expect that $k^{\ast}$ should not increase with $L$. The data are consistent with the statement.

\begin{figure}
\includegraphics[width=\columnwidth]{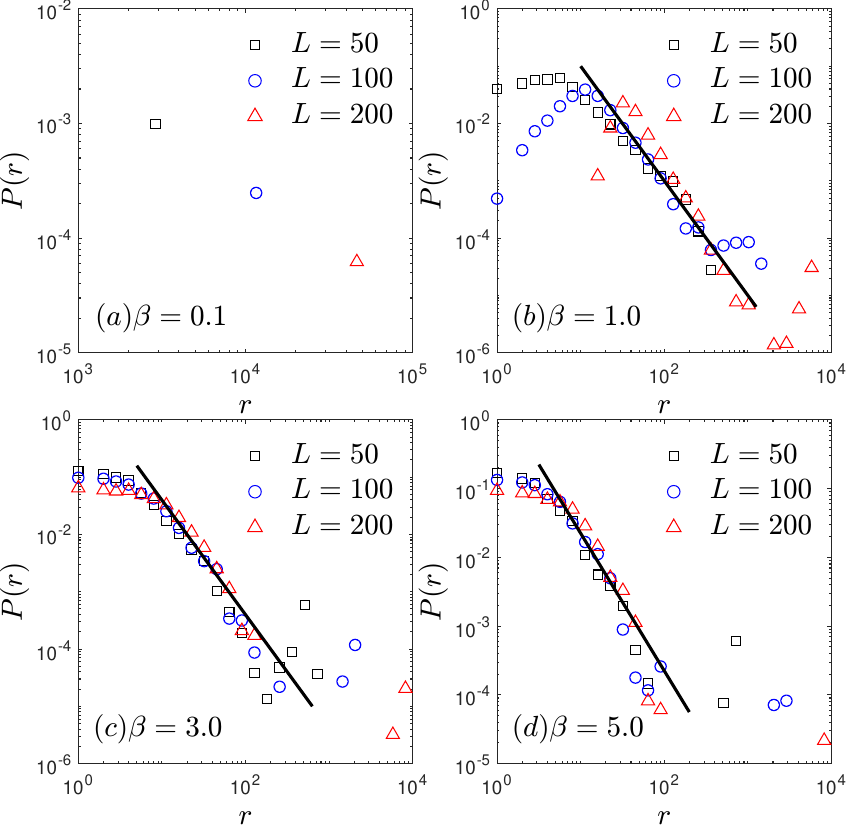}
\caption{The size distribution, $P(r)$, of record-breaking avalanches, $r$, for $\beta=$ (a) 0.1, (b) 1.0, (c) 3.0, and (d) 5.0, each for different system sizes. The solid line in (b)--(c) corresponds to the power law $P(r) \sim r^{-2}$. When $\beta=0.1$, the first avalanche is the largest and hence the data is insufficient for $P(r)$ to be quantitatively characterized.    }
\label{fig3}
\end{figure}

While the lifetime statistics of records depend on disorder and system size and could be a possible tool for prediction,  we now examine whether the distribution of record sizes also depends on extent of disorder. Let $P(r)$ denote the probability that a record has size $r$. In the fiber bundle model, this distribution was found to be independent of extent of disorder and distributed as $P(r) \sim r^{-1}$~\cite{kadar2020record}. We find that for RSNM $P(r)$ is independent of $\beta$ and $L$, and is power law distributed with $P(r) \sim r^{-\zeta_r}$ with $\zeta_r\approx2.0$, as shown in Fig.~\ref{fig3} for $\beta=1.0, 3.0, 5.0$. For $\beta=0.1$, the first avalanche is the largest and is thus the only record, hence  the avalanche size distribution consists of a single point. 

For $\beta=1.0, 3.0$, we find the existence of $k^{\ast}$ where average lifetime of records  is a maximum (see Fig.~\ref{fig2}). This feature is also seen for individual realizations, as can be seen from the inset of Fig.~\ref{fig1}(b)--(c). This makes sample wise prediction of fracture possible. Once the rank of records crosses $k^*$, the system can be said to accelerate towards failure. 

The presence of $k^{\ast}$ for only some $\beta$ can be rationalized as follows. The fracture of moderate disorder and smaller system sizes is avalanche dominated~\cite{shekhawat2013damage,moreira2012fracturing}. It would appear that in this regime, there is a non-trivial correlation between the failure strains and the strain at $k^\ast$, $\epsilon_{k^{\ast}}$. For lower disorder or larger system size, the fracture is nucleation dominated~\cite{shekhawat2013damage,moreira2012fracturing}. In the nucleation dominated regime, the failure is abrupt, making prediction difficult. This is consistent with our findings for the correlation between  failure strains and $\epsilon_{k^{\ast}}$ for low disorder and larger system  size (see below).

\begin{figure}
\includegraphics[width=\columnwidth]{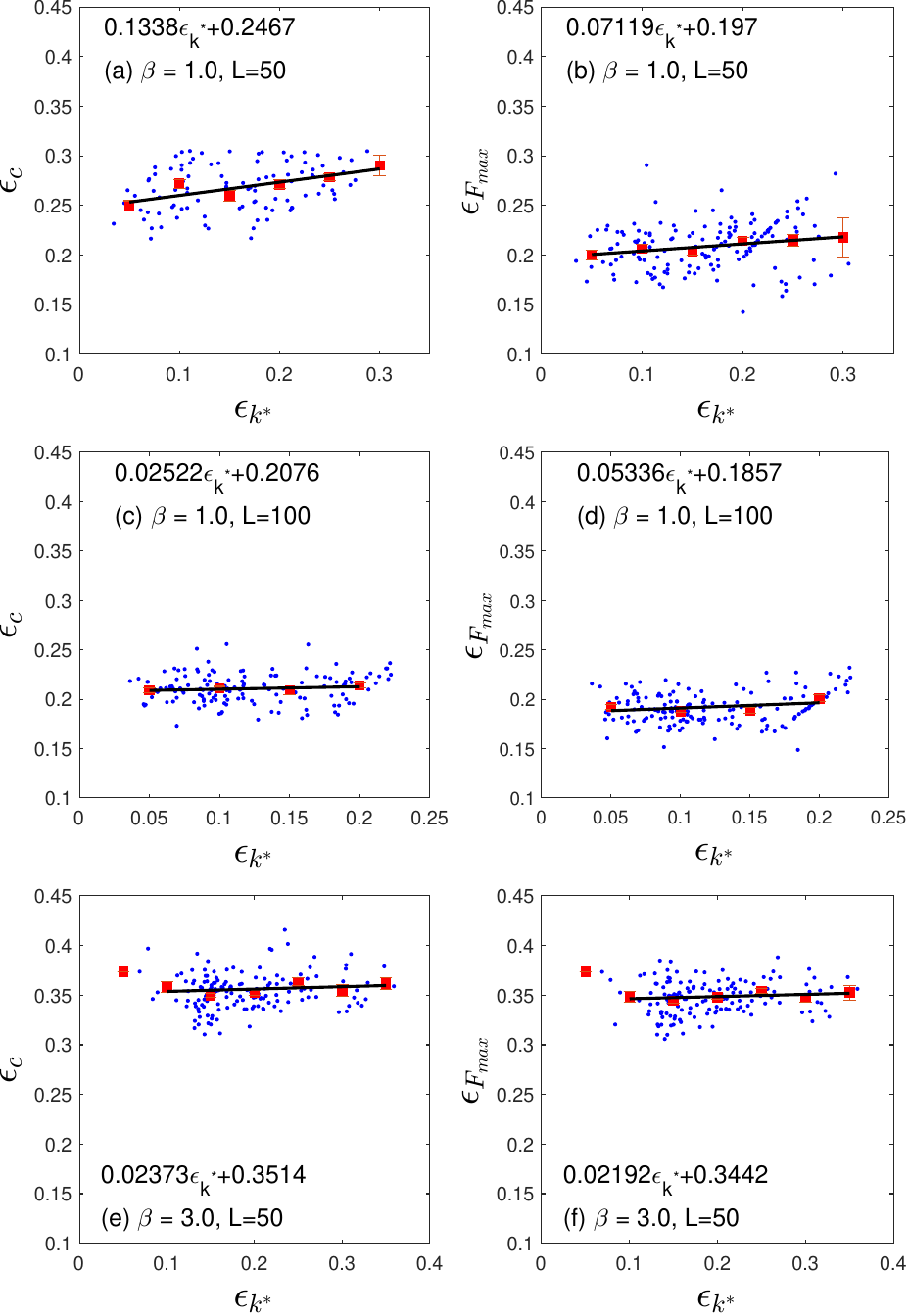}
\caption{Variation of $\epsilon_c$, the strain at final failure, and $\epsilon_{F_{max}}$, the strain at peak force, with  $\epsilon_{k^{\ast}}$, the strain $k=k^{\ast}$ for $\beta=1.0, 3.0$ as well as for $L=50, 100$. The straight lines are linear fits to the binned data, shown as red squares. As $L$ or $\beta$ increases, the slope decreases.
}
\label{fig5}
\end{figure}

We now ask whether this feature of the waiting times can be used to predict the failure strain.  Knowing $\epsilon_{k^{\ast}}$, we ask whether we can  predict $\epsilon_c$, the strain at which the whole network fractures across its width and $\epsilon_{F_{max}}$, the strain at peak force. The correlation between is $\epsilon_c$ and $\epsilon_{k^{\ast}}$, and $\epsilon_{F_{max}}$ and $\epsilon_{k^{\ast}}$ is shown in Fig.~\ref{fig5}(a, c, e) and Fig.~\ref{fig5}(b, d, f) for each of the realizations, as well as the binned data. We observe a linear relation between these pairs, as shown in Fig.~\ref{fig5}.
If the slope is zero, then the quantitative prediction of   $\epsilon_c$ and $\epsilon_{F_{max}}$, given $\epsilon_{k^{\ast}}$, is trivial, in the sense that they are independent of the knowledge of $\epsilon_{k^{\ast}}$. On the other hand, a non-zero slope leads to a non-trivial prediction of the failure strains, once $\epsilon_{k^{\ast}}$ is known. We observe that such a linear correlation exists for moderate disorder ($\beta=1.0$) [see Fig.~\ref{fig5}(a)-(d)], though the slope decreases with increasing system size.  Thus, we can predict the critical strain values as soon as we reach the maximum value of waiting time of records.
The decrease in slope with increasing system size is plausibly due to the fracture becoming more nucleation-type. For smaller disorder ($\beta=3.0$), the slope, while positive, is quite close to zero.

\begin{figure}
\includegraphics[width=0.8\columnwidth]{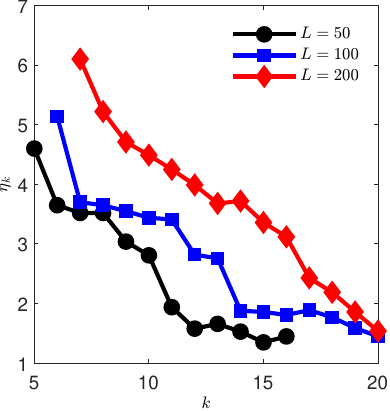}
\caption{The avalanche exponent $\eta_k$ for all avalanches that occur between the $(k-1)$-th and $k$-th records, for different system sizes. The data are for disorder characterized by $\beta=1$. }
\label{fig:dynamical-exponent}
\end{figure}

One of the earlier attempts at predicting fracture was to examine the strain dependent power-law exponent of the avalanche size distribution, without reference to records. The power-law exponent changed from a large value to a smaller one at strains closer to final fracture. We now ask how such a crossover is related to the record-related predictor, the maximum of the waiting times, $k^{\ast}$. To do so, we now calculate the effective strain (or time) dependent exponent of the avalanche distribution. Consider all avalanches between the $(k-1)$-th and $k$-th records. We find that the sizes of these avalanches are distributed as a power-law, whose effective exponent is denoted by $\eta_k$. We focus on $\beta=1.0$ for which the waiting time shows a prominent peak. The variation of $\eta_k$ with $k$ is shown in Fig.~\ref{fig:dynamical-exponent} for different system sizes.  It shows a decrease from a value close to $3.5$ to $\sim 1.7$. The crossover $k$ increases with $L$ and also becomes less sharp. This is in contrast to $k^{\ast} \approx 8$ for all $L$. Consequently, it seems plausible that by using record statistics, one could anticipate the occurrence of fracture earlier than would be indicated by the crossover time of avalanche distribution changing from off-critical to critical.

\section{Conclusions and discussion}

The statistical properties of fracture in disordered media represent an intriguing theoretical problem with important practical applications. One interesting aspect in this regard is the fact that heterogeneous materials do not break at once when subjected to external mechanical stress.  It has been experimentally observed that the response (acoustic emission) to increasing external stress takes place in discrete avalanches distributed over a wide range of scales. One of the important questions in fracture is whether avalanche statistics can be used as a predictor to final failure.

In this paper, we have simulated fracture of a random spring network model with different extent of disorder and system sizes to study the sequence of record-breaking avalanches during the course of failure. The waiting time associated with the record events shows a non-monotonic behavior and peak at a certain rank ($k^\ast$) of the record events. Such a $k^\ast$ exists only when the fracture is avalanche dominated. For percolation-type failure, we find that there is only one record. For nucleation dominated fracture, the waiting times between records decrease monotonically. In the avalanche-dominated regime, we show that the failure strain increases linearly with the strain at the maximum waiting time, hence it is possible to have a quantitative prediction of the failure strain. We also show that the strain dependent avalanche exponent decreases with record rank, consistent with the avalanches becoming critical beyond $k^\ast$. 

Compared to the predictor based on crossover from off-critical avalanches  to critical avalanches, the use of records as a predictor has the advantage of it being real-time and not requiring any further post processing. Also, as can be seen from Fig.~\ref{fig:dynamical-exponent}, the record rank for the crossover from off-critical to critical avalanches increases with system size. This is contrast to $k^\ast$ being very weakly dependent on the system size. Thus, it seems plausible that by using record statistics, one could anticipate the occurrence of fracture earlier.


%

\end{document}